\documentclass[aps,prl,twocolumn,superscriptaddress,floatfix,tightenlines,showpacs,notitlepage]{revtex4-1}
\usepackage{graphicx}
\usepackage{bm}
\usepackage[normalem]{ulem}
\usepackage{amsfonts}
\usepackage{color}
\usepackage{amsmath}    % need for subequations
\usepackage{epsfig}
\usepackage{subfigure}  % use for side-by-side figures
\usepackage[breaklinks,colorlinks = true,linkcolor = red,urlcolor=blue,citecolor=red]{hyperref}
\usepackage{amssymb}

\newcommand{\icts}{International Centre for
  Theoretical Sciences, Tata Institute of Fundamental Research,
  Bangalore 560089, India}
\newcommand{\iit}{Department of Chemical Engineering, Indian Institute of Technology Bombay, Mumbai 400076, India}
\begin{document}
\title{Sedimenting Elastic Filaments in Turbulent Flows}
\author{Rahul K. Singh}
\email{rahul.singh@icts.res.in}
\affiliation{\icts}
\author{Jason R. Picardo}
\affiliation{\iit}
\author{Samriddhi Sankar Ray}
\affiliation{\icts}

%\date{\today}
\begin{abstract}

We investigate the gravitational settling of a long, model elastic filament in
	homogeneous isotropic turbulence. We show that the flow produces a
	strongly fluctuating settling velocity, whose mean is moderately
	enhanced over the still-fluid terminal velocity, and whose variance has
	a power-law dependence on the filament’s weight but is surprisingly
	unaffected by its elasticity. In contrast, the tumbling of the filament
	is shown to be closely coupled to its stretching, and manifests as a
	Poisson process with a tumbling time that increases with the elastic
	relaxation time of the filament.

\end{abstract}

\maketitle

Sediments in turbulent flows are commonplace in nature and
industry.  Even so, theoretical progress in understanding how
objects settle under gravity while being buffeted by a turbulent carrier flow is recent.  Even in the
simplest case of tiny rigid spherical particles, the
interaction of these forces is rather intricate and leads to an enhancement of 
the settling velocity over its value in a still fluid---a fact that has been 
theoretically understood and quantified over just the past 
decade or so~\cite{M87,WM93,aliseda,ghosh,ayala2008effects,davila2001settling,Bec_gravity}.
More recent work has addressed the issue of
how spheroidal particles, with additional rotational degrees of freedom, orient themselves while settling~\cite{PumirPRL,AnandPRL}. 
A common feature of these prior studies is that they are
limited to rigid particles whose sizes, regardless of their anisotropy, are much
smaller than the dissipative Kolmogorov scale $\eta$ of the flow. These
constraints are certainly meaningful for a wide variety of applications, such as 
understanding the initiation of rain in warm clouds~\cite{Yau,Falkovich,Shaw,Bodenschatz970,Monchaux_pref_conc} and the radiative properties of cold clouds~\cite{baran,gallagher,pandit,voth}, which involve, respectively, turbulent suspensions of
 spherical water droplets and non-spherical ice crystals.
 
However, sediments that are both deformable and larger than the Kolmogorov
scale are just as ubiquitous. One important example is the sedimentation of
long deformable filaments, wherein flow-induced deformation modifies the net
drag force experienced by the filament ~\cite{Shelley}, thus coupling the
dynamics of conformation to settling.  This problem has been recently addressed
for low Reynolds number, non-turbulent flows~\cite{Marchetti}. However,
understanding the gravitational settling of filaments when the carrier flow is
turbulent (as encountered in papermaking~\citep{Lundell2011}
 and the fields of marine ecology
and pollution~\cite{Barnes,Lusher2015,Stelfox2016,Lebreton2018})
remains an important open problem.

In this paper, we address this issue through a combination of scaling analysis and
detailed numerical simulations on a model elastic filament in a homogeneous 
and isotropic turbulent flow. In particular, we show that the turbulent flow produces strong fluctuations in the settling velocity, while moderately enhancing its mean value over the terminal value in a still-fluid. We theoretically derive how the normalised variance
of the settling velocity scales with the
elasticity and inertia (mass) of the filament, as well as with the relative strengths of the accelerations due to gravity and turbulence. Our estimates are then verified
through detailed numerical simulations. Furthermore, borrowing ideas from 
the persistence problem of non-equilibrium statistical physics, 
we uncover a close connection between the two internal motions 
of tumbling and stretching, which accompany the unsteady yet inevitable
descent of the filament.

A simple model of a long filament, which retains enough internal structure to
exhibit both elasticity and inertia, is a chain of heavy inertial particles
connected through elastic springs~\cite{Bird}. Such chains are a useful
framework to understand the intricate interplay between elasticity and
turbulent mixing~\cite{PicardoPRL,Singh2020,RoyalChains} and provide insights which complement 
those obtained from other models of fibres~\cite{Brouzet_polymer,Verhille_3dconf_fiber,Brandt_fiber,Bec_Chain}.  In this context, the
most striking feature of filaments such as those we study is the manner in which they preferentially
sample the geometry of a turbulent flow: in the inertia-less (neutrally
buoyant) limit, such chains preferentially sample the vortical regions of the
flow, in both two and three dimensions (3D) though for different reasons
~\cite{PicardoPRL,RoyalChains}. This prediction, for the case of rigid chains
in 3D, was recently confirmed experimentally~\citep{Oehmke2020}. The
introduction of inertia (without gravity), counter-acts this tendency due to
centrifugal expulsion from vortices and decorrelation from the flow, resulting
in rather distinct dynamics~\cite{Singh2020}.  Here, we account for
gravitational acceleration and investigate the settling dynamics of these
\textit{elasto-inertial} chains.

We model a filament of mass $M$ as a chain of $N_b$ spherical beads.
Considering $\rho$ to be the mass density of the filament, we distribute the
mass uniformly over the beads, each of radius $a \ll \eta$, so that $\frac{4}{3}\pi
a^3\rho N_b = M$. The beads are then characterised
by a Stokesian relaxation time $\tau_p = \frac{2\rho a^2}{9\rho_f\nu}$, where
$\rho_f$ and $\nu$ are the density and kinematic viscosity of the carrier
fluid. Each bead, positioned instantaneously at ${\bf x}_j$, is connected to its nearest neighbours through (phantom) elastic links with which we associate a relaxation time scale $\tau_E$ (yielding an effective elastic
time scale ${\tau_E N_b(N_b+1)}/{6}$ for the filament~\cite{Collins2007,PicardoPRL}), thus rendering our
elasto-inertial chains extensible (and fully flexible). The dynamics of these model
filaments are then completely determined by the coupled equations of motion for the inter-bead
separation vectors ${\bf r}_j = {\bf x}_{j+1} - \textbf{x}_{j}$ and the
center-of-mass ${\bf x}_c$: 
	\begin{align} 
	\tau_p\ddot{{\bf r}}_j &= \left[{\bf u}({\bf	x}_{j+1},t)-{\bf u}({\bf x}_j,t) - \dot{{\bf r}}_j\right] + A\left [{\bm \xi}_{j+1}(t)-{\bm
		\xi}_j(t)\right ]  \nonumber\\
	&\hspace{2em}+\frac{1}{4\tau_E}\left (f_{j-1}{\bf
		r}_{j-1}-2f_{j}{\bf r}_{j}+f_{j+1}{\bf r}_{j+1}\right ) \nonumber\\
		&\hspace{8em}  \label{eqr}\\
		 \tau_p\ddot{{\bf x}}_c &=
		\left(\frac{1}{N_b}\sum_{j=1}^{N_b}{{\bf u}({\bf
		x}_j,t)}-\dot{{\bf
		x}}_c\right)+\frac{A}{N_b}\sum_{j=1}^{N_b}{\bm
		\xi}_j(t) - \tau_pg{\bf \hat z}.  \label{eqcm} 
		\end{align}
 Here, we use the FENE (finitely extensible
nonlinear elastic) interaction $f_j=(1-|\textbf{r}^2_j|/r^2_m)^{-1}$, with a
prescribed maximum inter-bead length $r_m$, to model the springs.
Independent white noises ${\bm \xi}_j(t)$, with a coefficient $A^2 = \frac{r^2_0}{6\tau_E}$, are imposed on the beads to set the equilibrium 
length $r_0$ for each segment of our filament (in the absence of flow). 
We choose a coordinate system such that the
acceleration due to gravity $g$ acts along the negative $z$-axis; the resulting
 gravitational forces acting on the
beads cancel out from Eq.~\eqref{eqr} but still
have an implicit effect on the dynamics of segments, via their action on the motion 
of the center-of-mass (which in turn determines the velocities sampled by the beads). 
Finally, the time-scales $\tau_p$
and $\tau_E$, as well as the acceleration due to gravity $g$,
allow us to define non-dimensional numbers in terms of analogous
(small-scale) quantities of the carrier turbulent flow, namely the Kolomogorov
time $\tau_\eta \equiv \sqrt{\nu/\epsilon}$ and acceleration
$a_\eta \equiv \left ({\epsilon^3}/{\nu}\right )^{1/4}$, where $\epsilon$
is the mean energy dissipation rate of the flow. The dynamics of our filaments
are thus completely determined by the
Stokes number $St \equiv \tau_p/\tau_\eta$ (a measure
of the inertia), the Froude number $Fr \equiv {a_\eta}/{g}$ (a measure of
the force of gravity), and the Weissenberg number $Wi \equiv \frac{N_b(N_b+1)\tau_E}{6\tau_\eta}$ (a measure of elasticity).  

Note that this simple model disregards inter-bead, hydrodynamic and 
excluded volume interactions, as well as bending stiffness~\citep{RoyalChains},
in an effort to more clearly reveal the
 fundamental interplay between elasticity, inertia, 
 and gravitational and turbulent acceleration. It is important to point
out that other models of filament dynamics, particularly in the context of
turbulent transport, are available~\cite{Brouzet_polymer,Verhille_3dconf_fiber,Brandt_fiber,Bec_Chain}.
However, studies for low Reynolds number flow show~\cite{Marchetti} that the
bead-spring approach to filaments still remains an important
framework~\cite{FilamentPRL,Schlagberger,Llopis,Delmotte} because of the
limitations of models based on slender-body theory~\cite{cox_1970,Xu}.

In our direct numerical simulations, we solve
Eqs~\eqref{eqr}-\eqref{eqcm}, using a
second-order Runge-Kutta scheme, together and simultaneously with the
three-dimensional, incompressible Navier-Stokes equations, driven to a
statistically stationary state by using a constant energy injection scheme. To simulate the carrier flow, we use
a de-aliased pseudo-spectral algorithm, spatially discretize the
Navier-Stokes equations on a 2$\pi$ periodic cubic box with $N^3 = 512^3$
collocation points, and use a second-order slaved Adams-Bashforth scheme for
time integration~\cite{James17}. The fluid velocity $\bf{u}({\bf x},t)$ obtained on the
regular periodic grid is interpolated, using trilinear interpolation,
to obtain the velocity ${\bf u}({\bf x}_{j+1},t)$ at the bead positions.
We choose the coefficient of viscosity $\nu = 10^{-3}$ to obtain a Taylor-scale 
Reynolds number $Re_\lambda \approx 200$. We evolve an ensemble of $10^4$ filaments, each with an equilibrium length 
$r_0 \approx 47\eta$ and $N_b = 10$ beads (we have checked that our results remain qualitatively
unchanged if we use a fewer number of beads, $N_b = 5$), and consider various values of $Wi$, $St$ and $Fr$. 

We begin our study
by examining the fluctuating settling velocities $v_z \equiv \dot{{\bf
x}}_c\cdot {\bf \hat z}$ of the filaments. Let us first consider the mean value 
$\langle v_z \rangle$, where the angular brackets
denote an average over the ensemble of chains and over time in the statistically stationary state. From Eq.~\eqref{eqcm}, denoting $a_z \equiv \ddot{{\bf x}}_c\cdot {\bf \hat z}$ as the vertical 
acceleration in the $z$ direction, the assumption of a mean settling velocity implies, by definition, that
$\langle a_z \rangle = 0$. Furthermore, since the noises are independent and of zero-mean,  we obtain
$\langle v_z \rangle = \langle \bar{u}_z \rangle - \tau_p g$, where $\tau_p g$ is the terminal settling velocity in a still-fluid, and $\bar{u}_z \equiv \frac{1}{N_b}\sum_{j=1}^{N_b}u_z({\bf x}_j,t)$ 
is the average $z$-component of the fluid velocity field sampled by the filament. If an object uniformly samples the flow, as tracer particles do, then $\langle \bar{u}_z \rangle = 0$. However, this is typically not the case for non-tracers: Heavy inertial particles (with $St > 0$) are known to preferentially sample descending regions of the flow, leading to $\langle \bar{u}_z \rangle < 0$ and thus an enhanced mean settling velocity. The magnitude of this effect, which may be quantified through $\Delta_V = \langle v_z \rangle/\tau_p g-1 =- \langle \bar{u}_z \rangle/\tau_p g$,  varies non-monotonically with $St$ as explained in~\citep{Bec_gravity}. 

\begin{figure}
\includegraphics[width=1.0\columnwidth]{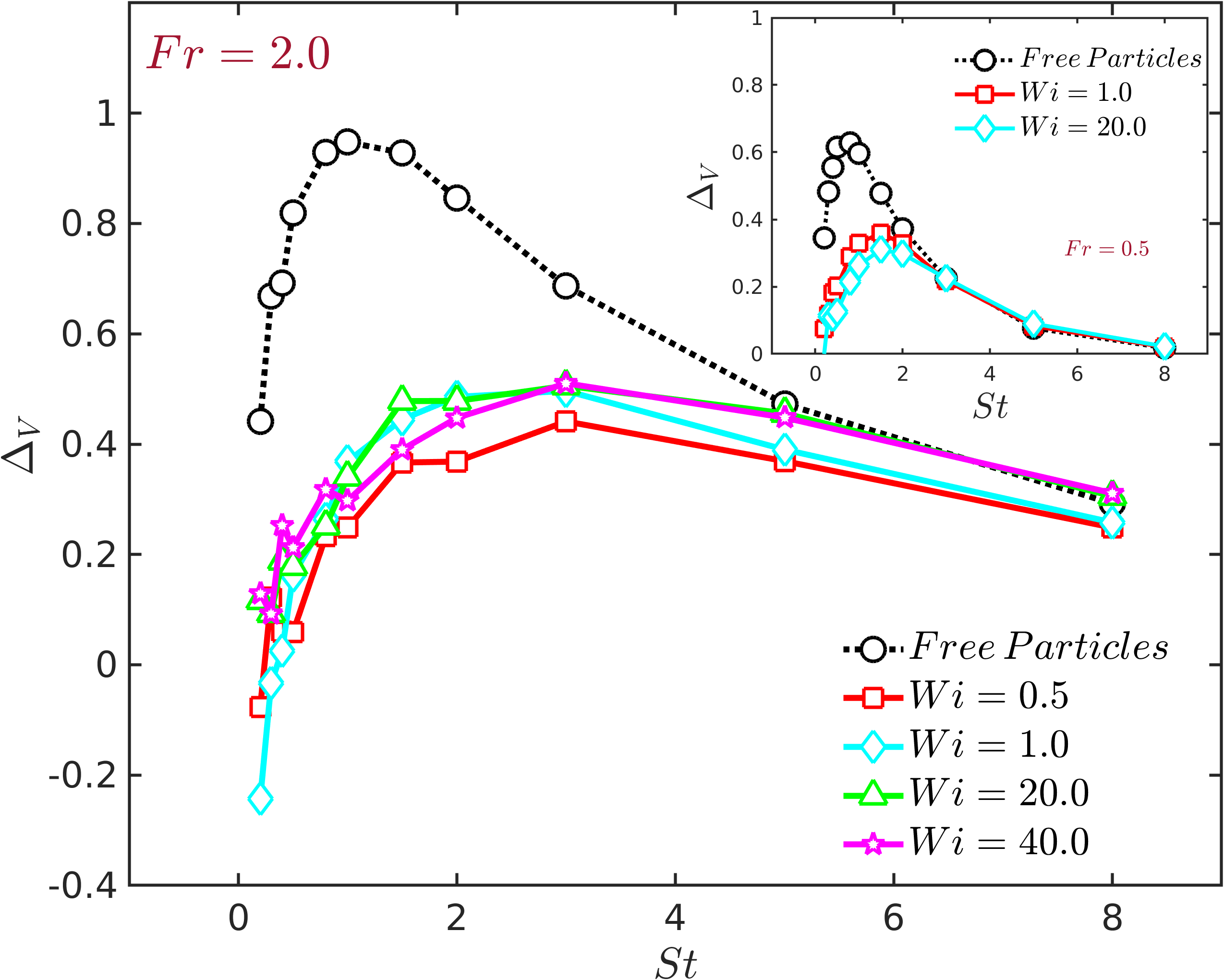}
	\caption{Plot of the relative enhancement in the mean settling velocity $\Delta_V$, over the still fluid value, as a function of $St$, for free,inertial, point particles and for filaments with different values of $Wi$. The inset shows results for a smaller $Fr$ than the main panel.}
\label{Delta_v}
\end{figure}

Now, as instantiated by our model, a filament can be thought of as a string of elastically-linked, inertial particles. So, how does this internal linking impact the mean settling velocity of filaments? Figure~\ref{Delta_v} answers this question by presenting simulation results of $\Delta_V$ as a function of $St$, for filaments with $Fr = 2$ (see the inset for $Fr = 0.5$) and various values of $Wi$, as well as for free inertial particles (whose dynamics are given by Eq.~\eqref{eqcm} with $A=0$ and $N_b = 1$). Clearly, the links counter-act the preferential sampling behaviour of inertial particles, resulting in a weaker enhancement of $\langle v_z \rangle$ for filaments. Indeed, at very small $St$ and for small to moderate $Wi$, $\Delta_V$ becomes negative and these filaments settle slightly \textit{slower} than they would in a still fluid. Figure~\ref{Delta_v} also suggests that the settling velocity is relatively insensitive to the elasticity of the filaments, a rather surprising observation which is reinforced below by considering the fluctuations about the mean value.

\begin{figure}
\includegraphics[width=1.0\columnwidth]{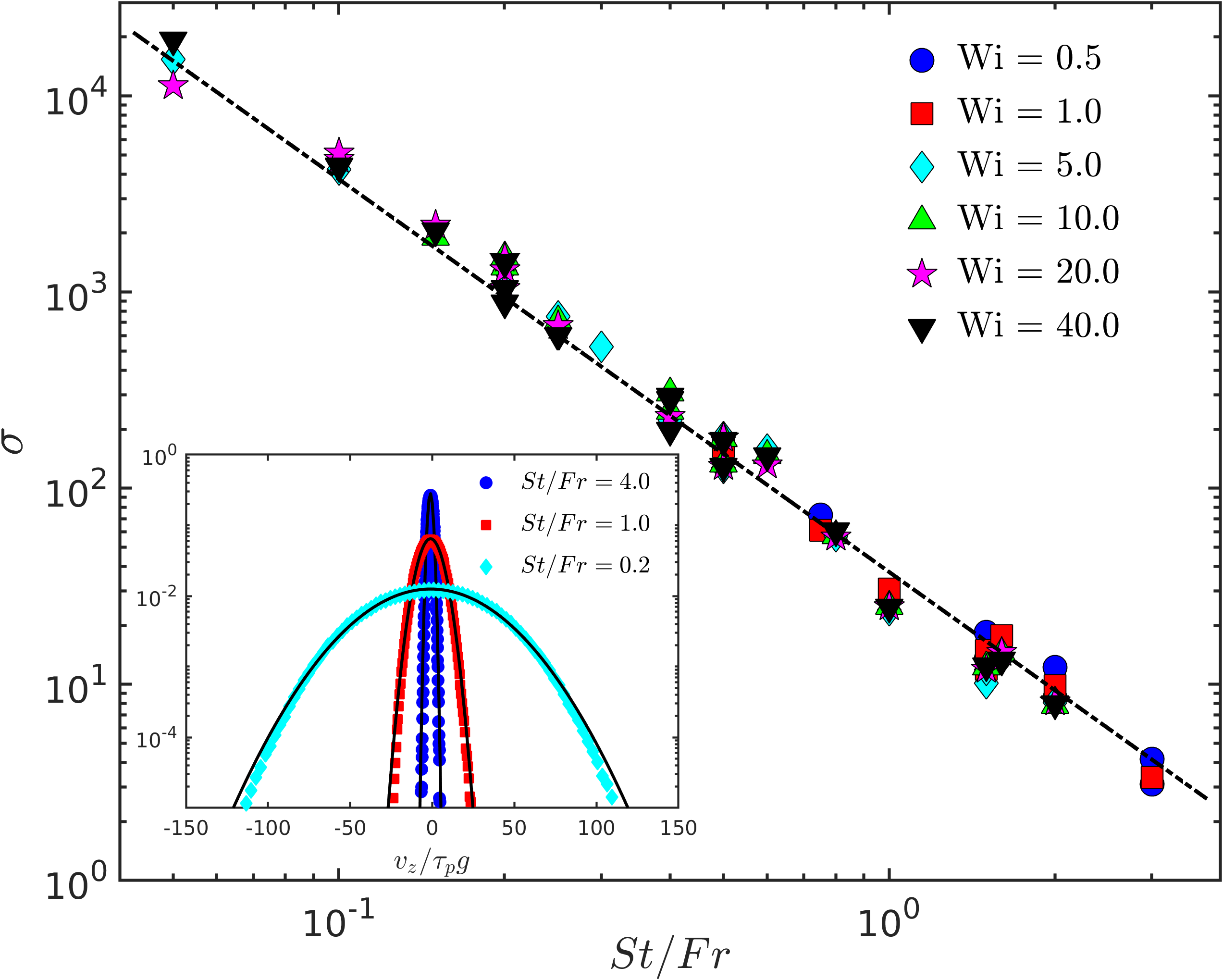}
	\caption{Loglog plot of the normalised variance $\sigma$ as a function of $St/Fr$; the different symbols correspond 
	to different values of $Wi$ (see legend) with the thick, black dashed line showing $(St/Fr)^{-2}$. In the inset, 
	we show representative plots of the pdfs (from which $\sigma$ is extracted) of the 
	settling speed $v_z$ rescaled by the corresponding $\tau_pg$  for $Wi = 40$ and different 
	values of $St/Fr$; the black lines are Gaussian fits.}
\label{pdf}
\end{figure}

The probability distribution function (pdf) of $v_z$, illustrated in the inset of Fig.~\ref{pdf}, shows that the turbulent flow can produce strong temporal fluctuations---much greater than $\Delta_V$---which become increasingly large and non-Gaussian (the black lines are Gaussian fits) as $St/Fr$ increases (the relevance of this ratio becomes clear below). The magnitude of these fluctuations is best quantified through the 
normalised variance $\sigma
\equiv \frac{\langle v^2_z\rangle}{\langle v_z \rangle^2} - 1$ of the
distribution. To obtain a theoretical estimate for this variance, we use Eq.~\eqref{eqcm} to calculate the second moment of the settling velocity $\langle v_z^2 \rangle$. 
Noting that $\sum_{j = 1}^{N_b} \sum_{k = 1}^{N_b} \langle \xi_{j,z}(t)\xi_{k,z}(t')\rangle = C\delta_{j,k}\delta(t-t')$, where the subscript $z$ 
denotes the $z$-components of the noise and $C$ is a constant (which absorbs the $N_b$ factor) with the dimension of inverse time, we obtain  
$\langle v_z^2 \rangle = \langle \bar{u}_z^2 \rangle + \tau_p^2 g^2 + \tau_p^2\langle a_z^2 \rangle + CA^2$, which
leads to the non-dimensional, normalised variance
\begin{equation}
	\sigma = \frac{\langle \bar{u}_z^2 \rangle + CA^2}{(a_\eta \tau_\eta)^2} \left(\frac{St}{Fr}\right)^{-2}, 
\label{variance}
\end{equation}
where, to extract the leading order behaviour, 
we set $\langle a_z^2 \rangle$ to zero 
(the validity of this approximation is addressed later) and use $\Delta_V < 1$.

This result is further simplified by the observation, from our simulations, that $\langle \bar{u}_z^2 \rangle \approx (3/2)E$, where $E$ is the mean kinetic energy of the flow. Furthermore, the small additive contribution of the noise to the variance, $CA^2 \propto 1/Wi$, is not relevant to understanding the effect of the turbulent flow on settling. (Indeed, one could use a non-stochastic, fixed equilibrium length, via a spring force proportional to $f_j (r_j - r_0)$, without affecting the qualitative dynamics of the chains in flow~\cite{RoyalChains}.) It follows, therefore, that $\sigma \sim (St/Fr)^{-2}$; the variance only depends on a single \textit{settling factor} $St/Fr$, which is the ratio of the still-fluid terminal settling velocity $\tau_p g$ to the Kolmogorov-scale velocity of the turbulent flow $u_{\eta} = a_\eta \tau_\eta$. 

Motivated by this scaling result, we plot the value of $\sigma$ obtained from our simulations, carried out by varying $Wi$, $St$ and $Fr$ \textit{independently} over a range of values, against $St/Fr$ in Fig.~\ref{pdf}.  We see that the leading order behaviour of the variance matches the scaling prediction $\sigma \sim (St/Fr)^{-2}$ (dotted line) remarkably well, over a wide range of $Wi$. Clearly, the extent of elasticity and therefore the stretching of the filament does not appreciably impact the settling velocity statistics. 

\begin{figure}
\includegraphics[width=1.0\columnwidth]{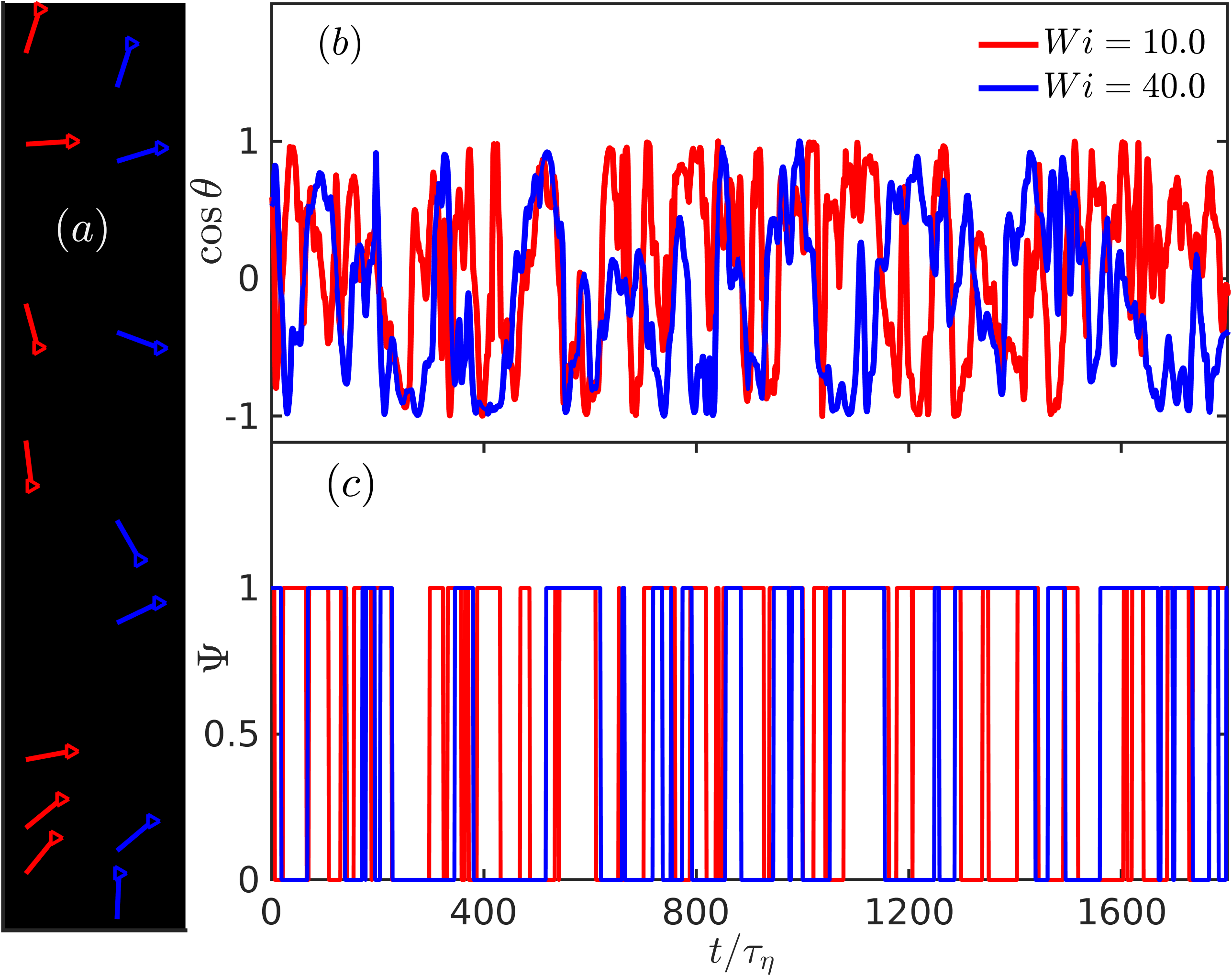}
	\caption{(a) Representative snapshots of the end-to-end vectors ${\bf R}$ of two filaments with $Wi = 10$ (red) and 40 (blue) 
	at different times (increasing downwards) as they sediment ($St =2,\; Fr = 0.5$); the vectors are projected on a two-dimensional plane 
	to illustrate the change in orientation $\theta$ with respect to the direction of gravity $-{\bf \hat z}$. Plots of 
	(b) $\cos\theta$ and (c) $\Psi$ (see text) of the same filaments 
	as a function of non-dimensional (with $\tau_\eta$) time: $\Psi = 1 (0)$ corresponds to an ``up'' (``down'') state of our filaments.}
\label{costheta}
\end{figure}

Before proceeding, it is useful to revisit the
approximation $\langle a_z^2 \rangle/g^2 \approx 0$ which lead to Eq.~\eqref{variance}. While our simulation data does not allow us direct access to this quantity, we can estimate its value by calculating the residual obtained on substituting our data into Eq.~\eqref{variance}. We find (not shown) that $\langle a_z^2 \rangle/g^2$ is practically independent of $Wi$, and although it has non-negligible values they are not large enough to disrupt the leading-order scaling with $St/Fr$, as is evident in Fig.~\ref{pdf}.

So far, we have only considered the settling of the filament as a whole. 
However, unlike for instance spherical
particles, these filaments have additional internal degrees of freedom which raise new 
questions, of which perhaps the most interesting is to understand how the filaments 
tumble as they descend through the turbulent flow. It is useful to recall that the tumbling of individual polymers---small elastic chains unaffected by inertia or gravity---has been studied in the context of simple shear flows~\cite{Steinberg-Tumbling,Celani_2005}. 

To study tumbling quantitatively, we consider the dynamics of the end-to-end
vector ${\bf R} \equiv \sum_{j = 1}^{N_b -1} {\bf r}_j$.  In
Fig.~\ref{costheta}(a), we show typical time traces of the end-to-end vector projected on a
two-dimensional plane, for two sedimenting
filaments with different $Wi$ (and $St = 2$, $Fr = 0.5$). Clearly the filaments
undergo complicated \textit{rotational} dynamics accompanied by
tumbling events. This behaviour is quantified through the cosine of the angle made by
${\bf R}$ with the $z$-axis, via  ${\bf R}\cdot{\bf \hat z} = R\cos \theta$, where $R=|{\bf
R}|$. In Fig.~\ref{costheta}(b), we show plots of the
time-series of $\cos \theta$ for the two filaments shown in panel (a). This
time series illustrates the seemingly continuous changes in the orientation of
the settling filaments with a suggestion that less elastic filaments tumble more frequently. Such observations 
naturally lead us to (a) suitably define the \textit{state} $\Psi$ of the filament as being either ``up'' or ``down'' and 
(b) to characterise the transitions between these two states.

\begin{figure}
\includegraphics[width=1.0\columnwidth]{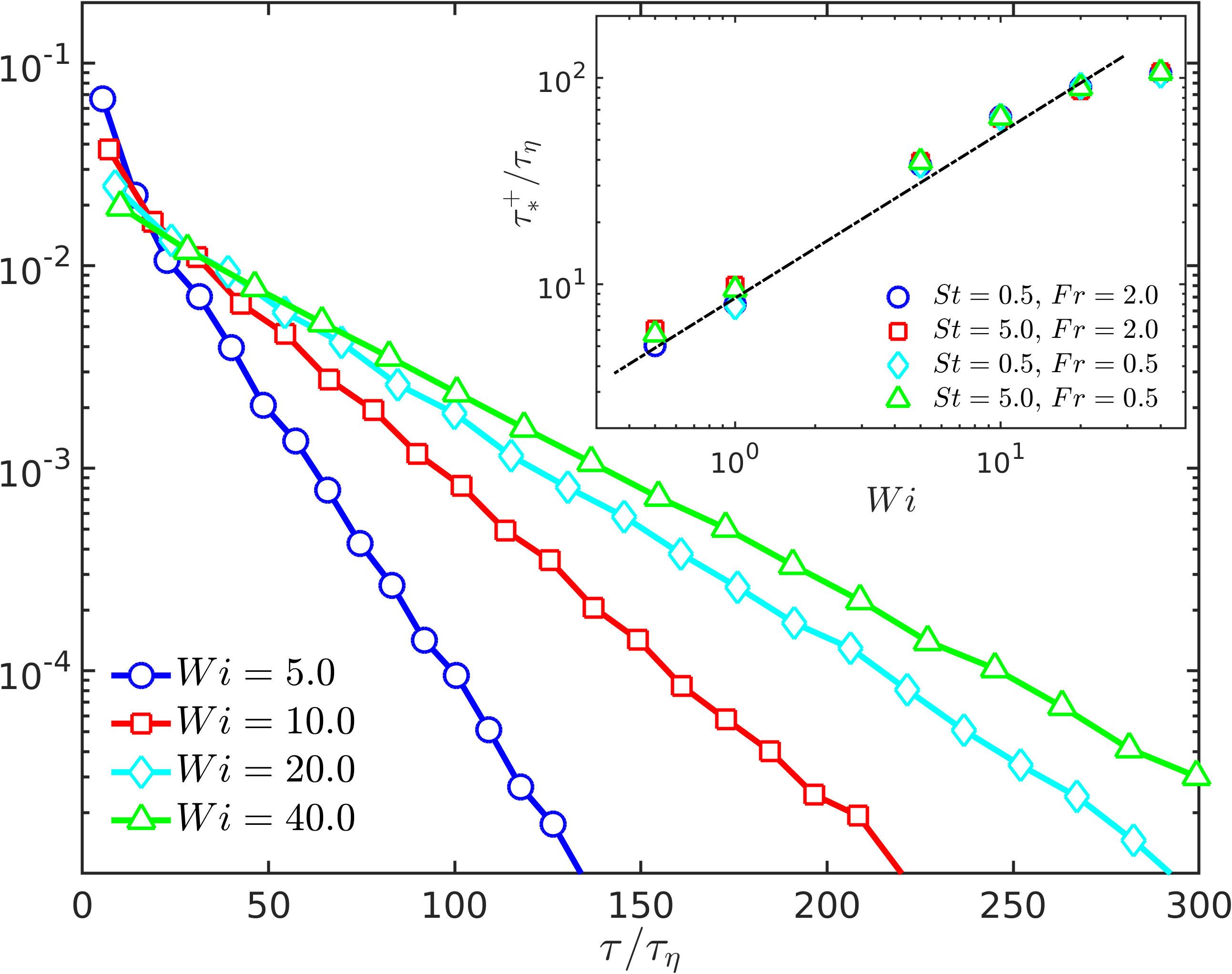}
	\caption{Probability distribution functions $P^+(\tau)$ of the residence time in the up ($\Psi = 1$) state for filaments with $St=Fr = 1.0$ and 
	different values of $Wi$. (Inset) Log-log plot of the characteristic time-scale $\tau_*^+$ versus $Wi$ for different values of $St$ and $Fr$; the thick black line indicates a scaling of $Wi^{4/5}$.}
\label{persistence}
\end{figure}

We define the up and 
down states as $\Psi = 1$ for $\cos\theta \geq 0$ and $\Psi = 0$ for $\cos\theta
< 0$, respectively. The apparently random
switching between the two states, clearly illustrated in Fig~\ref{costheta}(c), is quantified 
by calculating the pdfs $P^+ (\tau)$ ($P^- (\tau)$) of the residence time 
$\tau$ over which the filaments remain up (down). These distributions 
yield the probability of a filament in an 
up (or down) state to continue to remain in the same state for a duration of $\tau$.
We recall that questions of this sort---the so-called persistence problems---have a
special importance in areas of non-equilibrium statistical physics~\cite{Satya,Derrida,Satya-Review,Bray} and have,
more recently, been adapted to understand the geometrical aspects of turbulent
flows~\cite{Prasad,Bos,Akshay}.

In Fig.~\ref{persistence}, we show semi-log plots of $P^+(\tau)$
for filaments with $St=Fr = 1.0$, but different degrees of elasticity.
[We have confirmed, by varying the range of angles $\theta$ which define an up
or down state, that the precise definition of these states does not affect the results qualitatively;
furthermore, $P^+(\tau) = P^-(\tau)$ and hence
$\tau^+_* = \tau^-_*$, because the problem is symmetric to the transformation 
$r_j \rightarrow -r_j$ which reverses the end-to-end vector.] 
These distributions clearly show an exponential fall-off~\cite{Celani_2005}: $P^+(\tau) \sim \exp(-\tau/\tau^+_*)$,
which indicates that tumbling manifests as a Poisson process. The characteristic time scale 
$\tau^+_*$ has a weak dependence on $St$ and $Fr$, but increases systematically (see Ref.~\cite{Celani_2005} for single 
polymers in a shear flow) with $Wi$, as seen in the inset of Fig.~\ref{persistence}.

\begin{figure}
\includegraphics[width=1.0\columnwidth]{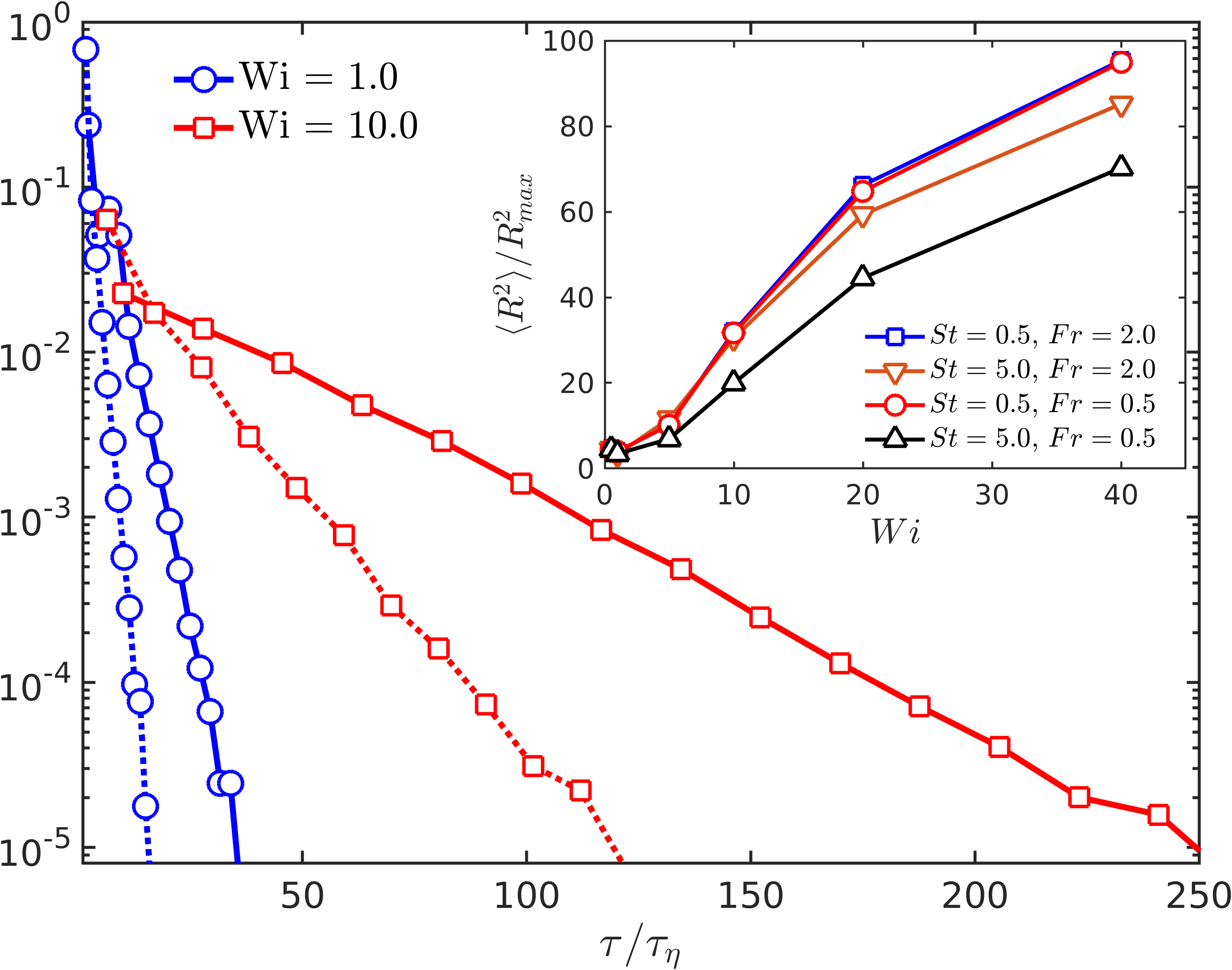}
	\caption{Conditioned probability distribution functions of time spent in the up ($\Psi = 1$) state, for filaments with relatively small ($\langle R \rangle_+ < \ell_1$; dashed lines) and large ($\langle R \rangle_+ > \ell_2$; solid lines) time-averaged lengths. The values of the thresholds for the two cases are $\ell_1 = 9\eta$, $\ell_2 = 20\eta$ for $Wi = 1$, and $\ell_1 = 40\eta$, $\ell_2 = 65\eta$ for $Wi = 10$. In both cases, $St = Fr = 1.0$.  (Inset) Plot of normalised $\langle { R^2}\rangle $, versus $Wi$ for different values of $St$ and $Fr$.}
\label{conditioned_pdfs}
\end{figure}

Why do more extensible filaments require longer times to tumble? The answer lies in the connection between the dynamic length of a filament and its tumbling. We expect a highly stretched, long filament
that spans multiple flow eddies to have a lower probability of experiencing the sequence of coordinated
drag forces required to cause a transition in its orientation. To test this hypothesis, we calculate the average end-to-end length of a filament over each interval of time spent in the up state $\langle R \rangle_+$. When a transition occurs, we record $\langle R \rangle_+$ along with the persistence time $\tau$, which then allows us to obtain the pdf of $\tau$ conditioned on the time-averaged length of the filament.  Fig~\ref{conditioned_pdfs} shows these conditioned pdfs for relatively short (dashed line) and long (solid line) filaments, defined as those with $\langle R \rangle_+ < \ell_1$ and $\langle R \rangle_+ > \ell_2$ ($\ell_1 <\ell_2$), where the values of the thresholds depend on $Wi$ and are given in the figure caption (small variations in these values do not affect our conclusions). Two values of $Wi$ are considered, and in both cases we see that when filaments are more stretched they do indeed take a longer time to tumble. (Note that a similar inverse relationship between length and tumbling rate has been recently observed in experiments on rigid, neutrally buoyant fibres in isotropic turbulence~\citep{Oehmke2020}.) Now, as $Wi$ increases, the filaments become more extensible and the distribution of $R$ broadens. This is demonstrated by the inset of Fig~\ref{conditioned_pdfs}, which presents the variation of $\langle R^2 \rangle$ with $Wi$, for all chains in the ensemble over all time. As a result, a larger $Wi$ filament is much more likely to be in a highly stretched state, which explains its tendency to persist in a given orientation for a longer time before tumbling. The consequent increase of $\tau_+$ with $Wi$, shown in the inset of Fig.~\ref{persistence}, appears to follow a power-law (the fitted dashed-line has an exponent of $4/5$) for small to moderate $Wi$, but then begins to level-off at large $Wi$, as the filament approaches its maximum length.

To summarise, we have analyzed two complementary aspects of the dynamics of 
long and heavy, elastic filaments in a turbulent flow: the fluctuating settling velocity, and the transitions in vertical orientation associated with tumbling.
For a given turbulent flow, we have found, rather surprisingly, that to leading order the 
weight of the filament only impacts its settling velocity (via the $St$ dependence of $\Delta_V$ and the $(St/Fr)^{-2}$ scaling of $\sigma$)
while the elasticity and consequent stretching of the filament only affects 
its tumbling. 

The Poisson distribution of tumbling times, uncovered by our persistence
analysis, brings to mind the Poissonian \textit{back-and-forth} motion of
filamentary, motile microorganisms. This tumbling-like motion is thought to be
an efficient strategy for foraging~\cite{Luchsinger,Stocker2635}, but
given that the Reynolds number of oceanic turbulence~\cite{Jimenez} is similar
to that in our study, it is tempting to consider in future works whether the
tumbling of marine filamentary microorganisms are a consequence of active
strategy or physical inevitability. Admittedly, these microorganisms are motile
and much smaller than our filaments; nevertheless, our results should motivate
work on longer marine organisms and their journey as they settle in the ocean.
Furthermore, this study is relevant to the sedimentation of passive marine pollutants, such as plastic debris from fishing gear~\citep{Stelfox2016,Lebreton2018}, as well as to fibre suspensions in industry~\citep{Lundell2011}.

\begin{acknowledgments} We are grateful to Dario Vincenzi for many 
	insightful discussions and for introducing us 
	to this problem. RKS and SSR also thank A. Kundu for useful discussions. 
	The simulations were performed on the ICTS
	clusters {\it Contra} and {\it Tetris} as well as the work stations from
	the project ECR/2015/000361: {\it Goopy} and {\it Bagha}. SSR and RKS
	acknowledge the support of the DAE, Govt.~of~India, under project
no.~12-R\&D-TFR-5.10-1100 and SSR is grateful to the DST (India) project MTR/2019/001553 for
financial help. JRP is thankful for support from the IIT-Bombay IRCC Seed Grant.
\end{acknowledgments}

\bibliography{ref_turb_chains}

\end{document}